\documentclass[twocolumn,showpacs,preprintnumbers,amsmath,amssymb]{revtex4}
\usepackage{graphicx}
\usepackage{dcolumn}
\usepackage{bm}

\begin{document}
\input epsf
\title{Velocity dependent interactions and a new sum rule in bcc He}
\author{Nir Gov}
\address{Department of Materials and Interfaces,
The Weizmann Institute of Science,\\
P.O.B. 26, Rehovot, Israel 76100}

\begin{abstract}
Recent neutron scattering experiments [PRL,{\bf 88},p.195301
(2002)] on solid $^4$He, discovered a new optic-like mode in the
bcc phase. This excitation was predicted by a recently proposed
model that describes the correlated atomic zero-point motion in
bcc Helium in terms of dynamic electric dipole moments.
Modulations of the relative phase of these dipoles between
different atoms describes the anomalously soft T$_1$(110) phonon
and two new optic-like modes, one of which was recently found in
the neutron scattering experiments. In this work we show that the
correlated dipolar interactions can be written as a velocity
dependent interaction. This then results in a modified f-sum rule
for the T$_1$(110) phonon, in good agreement with the recent
experimental data.
\end{abstract}

\maketitle

The phonons in a quantum solid, such as solid He, show various
quantum effects arising from the unusually large zero-point motion
of the atoms. The strong quantum pressure in solid He lowers the
density and introduces large anharmonic terms in the interatomic
potential. In bcc He the potential even has a double-well feature
\cite{glyde,niremil}. Describing the phonon spectra of this highly
anharmonic quantum solid in a consistent manner has therefore
presented a theoretical challenge over the years.

The quantum effects appear in the first moment of the dynamic
structure factor $S(\omega,k)$, as measured by neutron scattering
\cite{minki}. The Debye-Waller factor $d(k)$ is unusually small
due to the large spread in the atomic positions. In addition there
have been measurements \cite{minki} of apparent oscillations of
the Debye-Waller factor as a function of the scattering wavevector
$k$. These oscillations seemed to be confined to some modes, while
absent in the transverse T$_1$(110) mode. The explanation of these
unusual oscillations was given in terms of interference effects
between the single and multi-phonon excitations, whereby intensity
is transferred between them \cite{glyde}. The mixing of the phonon
modes is attributed to the anharmonicity of the interatomic
potential. From symmetry considerations these types of
calculations predict that the oscillatory contribution to $d(k)$
vanishes both at the Brillouin zone center and boundary (z.b.).

Recently there has been a new neutron scattering experiment
\cite{tuvyprl} in solid $^4$He. The main surprise in this
experiment was the discovery of a new, gapped excitation branch in
bcc $^4$He. This excitation was found at the energy predicted by a
new model of the dynamic effects of the zero-point motion on the
phonon spectra \cite{niremil,nirbcc}. Usually a bcc lattice is not
expected to support optic-like phonons, so the measurement of this
new excitation branch provides support for the basic idea of the
new model, namely the introduction of an additional degree of
freedom in bcc He. Within the new model, this degree of freedom is
the relative phase of a correlated part of the zero-point motion
between the atoms. The latest experiment has also provided data
which shows apparent oscillations of the Debye-Waller for the
transverse T$_1$(110) mode \cite{tuvythesis}. It is this data
which we shall analyze in this letter, in terms of the new model.

In previous papers \cite{niremil,nirbcc} we have presented a new
model to describe some of the correlations in the atomic
zero-point motion in bcc Helium. The main point is that there is
an anharmonic, low energy mode of atomic motion along the major
axes of the bcc lattice, which is not described by the harmonic
approximation. In bcc He this mode has energy $E_{0}\sim 3-10$K
\cite{niremil}, and is the only free parameter in our model, which
we take from empirical data. Additionally, the low mass of the He
atom allows relatively large breaking of the Born-Oppenheimer
approximation which results in motion-induced electric dipoles,
due to relative displacement of the nuclei with respect to the
electronic cloud \cite{niremil,niremilafm}. When these dipoles are
uncorrelated they introduce a negligible addition to the usual
Van-der Waals interaction. By contrast, correlated and highly
directional atomic motion results in dipolar interactions between
these dipoles, which is of the order of the energy of the local
atomic motion $E_{0}$. There can therefore arise a state of
quantum resonance between these interacting electric dipoles and
the atomic motion. A resonant state between the atomic motion and
electric dipolar interactions lowers the ground-state energy of
the system. In the ground-state of the bcc phase we found
\cite{niremil} that the dipolar interaction is minimized when the
dipoles have the phase relation shown in Fig.1.

The dipolar interactions between the motion-correlated dipoles is
\cite{niremil}
\begin{eqnarray}
E_{dip} &=&\sum_{i\neq 0}{\bf \mu }_{0}\cdot{\bf \mu
}_{i}\left[\frac{3
\left( {\hat {\bf \mu}}\cdot {\hat {\bf r}_{i0}}\right)^2 -1}{%
\left| {\bf r}_{i0}\right| ^{3}}\right] \nonumber \\
\label{edip}
\end{eqnarray}
where the electric dipole moment in bcc He turns out to be
\cite{niremil}: ${\bf \mu }\simeq e \cdot 0.01$, and ${\bf
r}_{i0}={\bf r}_{0}-{\bf r}_{i}$ and the index $i$ runs over all
atoms. The resonance condition means that $E_0$ is the frequency
of both the oscillating dipoles and the atomic motion, i.e.
$E_{dip}\equiv -E_0/2$ ($\simeq -3$K in bcc $^4$He). Note that the
dipole moments are dynamic, and the expectation value of the
dipole moment of the crystal and of each atom is zero, as required
by time reversal symmetry. It is the matrix element of correlated
dipoles $\left\langle {\bf \mu }_{0}{\bf \mu }_{i} \right\rangle$
which is non-zero, and appears in (\ref{edip}).

The modulations of the relative phases of the dipoles on different
atoms coincide with only one phonon mode, due to the lower
symmetry of the dipolar ground-state compared with the bcc lattice
(Fig. 1) \cite{niremil}. This mode turns out to be the anomalously
soft T$_{1}$(110) phonon, which agrees very well with the
experimental data \cite{niremil}. In addition our model predicts
two new modes which are gapped (optic-like) \cite{niremil,nirbcc}.
One of these modes was recently observed in neutron scattering
experiments \cite{tuvyprl}. Since the dipolar degree of freedom
involves atomic motion, it couples to the atomic density operator.
A flipping of a dipole with respect to the ground-state
configuration (Fig.1) results in a dynamic density fluctuation. A
flipping of two adjacent dipoles forms the localized excitation of
energy $2E_0$ \cite{niremil} recently found by neutron scattering
\cite{tuvyprl}. Our model therefore gives a good description of
some effects of the anharmonic potential in bcc He, as they are
manifested in the excitation spectrum.

Since the dipolar interactions arise from correlated atomic
motion, they can be naturally written as a velocity-dependent
potential. Such a term will modify the Nozieres-Pines f-sum rule,
which results from a Hamiltonian where the momentum operator
appears only in the kinetic energy term \cite{pines} (i.e. no
velocity-dependent interactions). In this paper we calculate the
f-sum rule for the T$_{1}$(110) phonon, which we describe as a
phase modulation of the correlated dipolar array \cite{niremil}.

Before proceeding let us make a short comment concerning the
appearance of velocity-dependent interactions in physics. One
example is superfluid $^4$He \cite{dalfovo}. In the superfluid a
velocity-dependent interaction is introduced into the energy
density functional in order to take into account short-range
atomic motion correlations. These are termed "backflow" and
introduce a non-local velocity-dependent term to the potential
energy. The result is a modified f-sum rule for the phonon-roton
dispersion in bulk superfluid $^4$He (Eq.(44) in \cite{dalfovo}).
All this is very similar to our results, only that we do not
introduce any unknown fitting parameters, as is done for the
superfluid case.

The dipolar interaction matrix element $X({\bf k})$ in the
presence of a phase modulation (i.e. a T$_{1}$(110) phonon), is
the Fourier transform of (\ref{edip}) \cite{niremil}
\begin{eqnarray}
X\left( {\bf k}\right) &=&\sum_{i\neq
0}{\bf \mu }_{0}\cdot{\bf \mu }_{i}\left[\frac{3
\left( {\bf {\hat \mu}}\cdot {\hat {\bf r}_{i0}}\right)^2 -1}{%
\left| {\bf r}_{i0}\right| ^{3}}\right] \nonumber \\
&&\  \ \times \exp \left[ 2\pi i{\bf k}\cdot {\bf r}_{i0} \right]
\label{xk}
\end{eqnarray}
shown in Fig.2.

By analogy with an oscillating electromagnetic field interacting
with an atom \cite{cohen}, we can write the electric dipole moment
in term of a local momentum operator
\begin{equation}
\tilde{\mu}=e\tilde{X}\rightarrow \tilde{P}\cdot \frac{e}{m
\omega_0} \label{xtop}
\end{equation}
where $\omega_0=E_{0}/\hbar$ is the resonance frequency of atomic
and dipolar vibration, and the $\tilde{}$ symbol is used for
operators.

The space representation of the dipolar interaction energy
(\ref{edip}) between the zero-point dipoles in the ground-state
can now be written in terms of the local momentum operators
\begin{eqnarray}
X&=&\sum_{i,j,i\neq j}\tilde{\mu}_{j}\cdot \tilde{\mu}_{i}\left[
\frac{3
\left( {\bf {\hat \mu}}\cdot {\hat {\bf r}_{ij}}\right)^2 -1}{%
\left| {\bf r}_{ij}\right| ^{3}}\right] \nonumber \\
&=&\left(\frac{e}{m \omega_0}\right)^2\sum_{i,j,i\neq j}\left[ \frac{3
\left( {\bf {\hat \mu}}\cdot {\hat {\bf r}_{ij}}\right)^2 -1}{%
\left| {\bf r}_{ij}\right| ^{3}}\right] \tilde{P}_{i}
\tilde{P}_{j} \label{xij}
\end{eqnarray}
where ${\bf r}_{ij}={\bf r}_{j}-{\bf r}_{i}$. Note that similar to
the treatment \cite{dalfovo} of superfluid $^4$He, we find a
non-local kinetic term.

The contribution of an interaction $X$ (\ref{xij}) to the f-sum
rule is found by performing the usual double commutator with the
density operator \cite{pines}
\begin{eqnarray}
\left\langle \left[ \left[\rho_{k},X\right],\rho_{-k}\right]
\right\rangle &=&\frac{\hbar^2 k^2}{m} \frac{e^2}{m \omega_0^2}
\sum_{i,j,i\neq j}\hat{\bf P}_{j}\cdot\hat{\bf P}_{i}\left[
\frac{3
\left( {\bf {\hat \mu}}\cdot {\hat {\bf r}_{ij}}\right)^2 -1}{%
\left| {\bf r}_{ij}\right| ^{3}} \right] \nonumber \\
&&\  \ \times \exp \left[ 2\pi i{\bf k}\cdot {\bf r}_{ij}\right]  \nonumber \\
&=&N\frac{\hbar^2 k^2}{m} \frac{|X(k)|}{E_0}
\label{commu}
\end{eqnarray}
where $\rho_{k}=\sum_{i}e^{-ik\cdot r_{i}}$ is the standard
density operator and $\hat{\bf P}$ is a unit vector of the local
momentum so that we keep track of the phase relation between the
correlated motion of different atoms. The commutator (\ref{commu})
is performed at equal times \cite{scat}. When calculating the
summation in (\ref{commu}) we have to specify the phase relation
between the dipoles on different sites. If there is no correlation
between the dipoles, the interaction $X(k)$ is zero, and the
correlation function of the atomic displacements described by
these dipoles, i.e. $S(\omega,k)$, is equally zero. In the bcc
phase we have the dipoles correlated in the ground-state as shown
in Fig.1, with resulting non-zero dipolar interactions (Fig.2).
The momentum operators in (\ref{xij}) replace the usual kinetic
energy term for the part of the atomic motion which is described
by the dipolar degree of freedom.

Using this double commutator (\ref{commu}) in the usual definition
of the dynamic structure-factor $ S(\omega,k)$
\cite{pines,kittel}, we get that the new sum rule for the
T$_{1}$(110) phonon is
\begin{equation}
M_{1}=\int_{0}^{\infty} d\omega \omega S(\omega,k)=N\frac{\hbar^2 k^2}{2m}\left( \frac{2|X(k)|}{E_0} \right)d(k)^2
\label{dipsum}
\end{equation}
where just as for usual phonons, the single-phonon contribution to
the f-sum rule is modified by the Debye-Waller factor
$d(k)=exp^{-\left\langle u^2 \right\rangle k^2}$ (ACB sum
rule)\cite{glyde,acb}, with $\left\langle u^2 \right\rangle$ being
the isotropic zero-point spread of the atomic wavefunction. We
assumed here that the dynamic structure factor $S(\omega,k)$, i.e.
the correlations in atomic displacements, for the T$_{1}$(110)
phonon are completely described by the dipolar degree of freedom
alone. This is justified by the excellent agreement between our
calculation of the and the T$_{1}$(110) spectrum experimental data
\cite{niremil}.

The new sum rule we find for the T$_{1}$(110) phonon
(\ref{dipsum}) is essentially the usual f-sum rule multiplied by
the function $2|X(k)|/E_0$. We find that at $k\rightarrow 0$, we
recover the usual f-sum rule, since $|X(k\rightarrow 0)|
\rightarrow E_0/2$. While the usual f-sum rule follows from
particle conservation, the number of excited dipoles out of the
correlated ground-state is not strictly conserved. In particular,
when the function $X(k)$ goes to zero at the Brillouin z.b., the
commutator (\ref{commu}) vanishes. The vanishing of $M_{1}(k)$ can
be understood as a vanishing of both the restoring force of the
dipolar interactions and therefore the dipole-induced correlations
between the atomic motion on different sites. Exciting a dipole
does not affect the others at this particular momentum, and the
neutron can not transfer any energy to this degree of freedom.
Note that $M_{1}\le N\hbar^2 k^2/2m$ since $2|X(k)|\le E_0$, so
that the energy constraint which specifies that the maximum energy
transferred by the neutron to the atom is the given by recoil
energy, is satisfied.

Before comparing to the experimental data we note that in the
experiments the neutron scattering is usually done in the second
or higher Brillouin zone, so the periodic behavior of $X(k)$
(Fig.2) has to be taken into account. The experimental set-up is
described elsewhere \cite{tuvyprl,minki}, and both sets of data
are for bcc $^4$He. Our calculation is equally applicable for bcc
$^3$He \cite{niremilafm}, for which there is at present no neutron
scattering data.

The calculated $M_{1}(k)$ (\ref{dipsum}) is shown in Fig.3
compared to the neutron scattering data \cite{tuvythesis,minki}.
For $d(k)$ we use the value $\left\langle u^2
\right\rangle=0.17$\AA$^2$, taken from the measured
\cite{greywall} Debye temperature of $\sim27$K. By plotting the
ratio $M_{1}(k)/\left([{\bf k}\cdot {\bf e}(k)]^2 d(k)^2\right)$
we can easily see deviations from the usual f-sum rule, which
gives a constant value (${\bf e}(k)$ being the phonon
polarization) \cite{glyde}. The data in Fig.3 is plotted against
the full scattering momentum of the T$_1$(100) phonon as measured
in the most recent experiments \cite{tuvyprl,tuvythesis}, so both
the older data \cite{minki} and the calculation are mapped
accordingly.

We find that our calculation is in good agreement with the
experimental data. The calculated oscillation in the intensity is
observed, in contrast with the usual f-sum rule result. The
observed periodicity is different from the one calculated
previously for interference terms \cite{glyde}, which gives a
correction that vanishes both at the z.b. and the zone center.
Additionally, the multi-phonon interference calculation does not
predict the complete vanishing of the intensity at any momentum,
unlike the result of the model presented here. Note that both our
model and previous calculations attribute the oscillations in the
scattered intensity to anharmonic effects.

The prediction that the intensity should vanish at the Brillouin
z.b. is not observed (Fig.3), only a marked reduction in the
scattering intensity. This can arise from the finite resolution of
the experiment \cite{tuvyprl} allowing in stray scattering from
other phonons with similar energy (at the z.b. the T$_1$(110)
branch meets other phonon branches). At the z.b. with the smaller
momentum ($k\simeq 1.85$\AA$^{-1}$) there is higher scattering
intensity then at the higher momentum z.b. ($k\rightarrow
2.85$\AA$^{-1}$), as expected for stray scattering from other
phonons. Future high resolution data at the zone boundaries is
needed to resolve this question.

In this work we presented a modified f-sum rule for the
anomalously soft T$_{1}$(110) phonon mode of bcc He. We find an
apparently oscillating Debye-Waller factor and a complete
vanishing of the first moment at the Brillioun zone boundary. The
new f-sum rule is qualitatively different from the standard
response of a harmonic phonon. This unusual result is due to
correlations in the atomic zero-point motion, which can be written
in terms of a velocity-dependent interaction. The calculated sum
rule is then compared with recent neutron scattering data, which
does show some of the predicted features. Combined with the recent
observation of a new excitation mode \cite{tuvyprl}, the sum rule
data presented here provides extra support for the proposed model
of correlated zero-point electric dipoles in the ground-state of
bcc He \cite{niremil,nirbcc}.

\begin{acknowledgments}
I thank Emil Polturak, Tuvy Markovitch and Emmanuel Farhi for
numerous discussions and access to their data. I also wish to
thank Gordon Baym and Tony Leggett for many useful discussions and
suggestions. This work was supported by the Fulbright Foreign
Scholarship grant and the Perlman Postdoctoral Fellowship. Support
was also provided by NSF grant no. DMR-99-86199 and NSF grant no.
PHY-98-00978 while in the University of Illinois at
Urbana-Champaign.
\end{acknowledgments}

\begin{figure}
\input epsf \centerline{\ \epsfysize 4.5cm \epsfbox{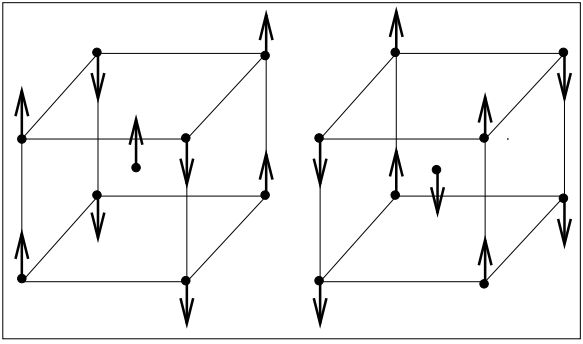}}
\caption{The two degenerate 'antiferroelectric' dipole
arrangements in the ground-state of the bcc phase. The system is
in quantum resonance between these two configurations with
resonance frequency $\omega_0=E_0/\hbar$. The arrows show the
instantaneous direction of the dipoles along one of the major
axes. Similar dipolar arrays exist along the orthogonal major
axes.}
\end{figure}

\begin{figure}
\input epsf \centerline{\ \epsfysize 6cm \epsfbox{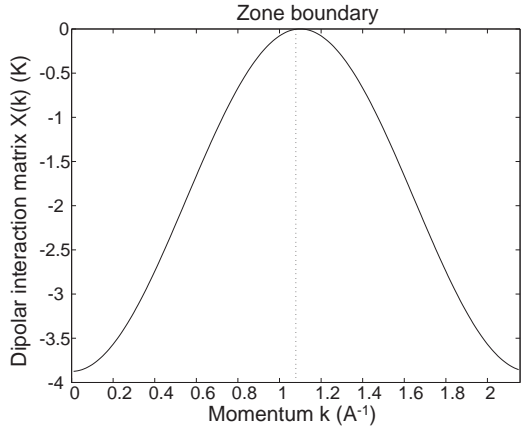}}
\caption{The dipolar interaction matrix element $X(k)$ (Eq.\ref{xk}) in the bcc phase of $^4$He, along the (110) direction.}
\end{figure}

\begin{figure}
\input epsf \centerline{\ \epsfysize 6.5cm \epsfbox{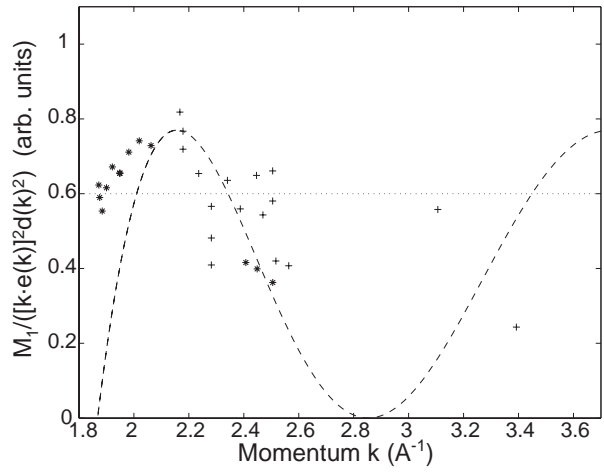}}
\caption{The ratio $M_{1}(k)/\left([{\bf k}\cdot {\bf e}(k)]^2
d(k)^2\right)$ given by Eq.\ref{dipsum} (dashed line), compared
with the experimental data for the T$_{1}$(110) phonon: crosses
\cite{minki} and stars \cite{tuvythesis}. The usual form for a
harmonic phonon is given here by the horizontal dotted line.}
\end{figure}

\end{document}